\documentclass[sigconf]{acmart}
\usepackage{svg}

\usepackage{tikz}
\usepackage{array,multirow,graphicx}
\usepackage{adjustbox}
\usepackage{amsmath}
\usepackage{bm}
\usepackage{colortbl}
\usepackage{enumitem}
\usepackage{tabu}
\usepackage{balance}
\usepackage{footnote}

\AtBeginDocument{%
  \providecommand\BibTeX{{%
    \normalfont B\kern-0.5em{\scshape i\kern-0.25em b}\kern-0.8em\TeX}}}

\setcopyright{acmcopyright}

\copyrightyear{2022} 
\acmYear{2022} 
\setcopyright{rightsretained} 
\acmConference[CIKM '22]{Proceedings of the 31st ACM International Conference on Information and Knowledge Management}{October 17--21, 2022}{Atlanta, GA, USA}
\acmBooktitle{Proceedings of the 31st ACM International Conference on Information and Knowledge Management (CIKM '22), October 17--21, 2022, Atlanta, GA, USA}
\acmDOI{10.1145/3511808.3557714}
\acmISBN{978-1-4503-9236-5/22/10}

\begin{document}

\title{TripJudge: A Relevance Judgement Test Collection for TripClick Health Retrieval}

\author{Sophia Althammer}
\email{sophia.althammer@tuwien.ac.at}
\affiliation{%
  \institution{TU Wien}
  \city{Vienna}
  \country{Austria}
}

\author{Sebastian Hofst{\"a}tter}
\email{sebastian.hofstaetter@tuwien.ac.at}
\affiliation{%
  \institution{TU Wien}
  \city{Vienna}
  \country{Austria}
}

\author{Suzan Verberne}
\email{s.verberne@liacs.leidenuniv.nl}
\affiliation{%
  \institution{Leiden University}
  \city{Leiden}
  \country{Netherlands}
}

\author{Allan Hanbury}
\email{hanbury@ifs.tuwien.ac.at}
\affiliation{%
 \institution{TU Wien}
 \city{Vienna}
 \country{Austria}}

\renewcommand{\shortauthors}{Althammer, et al.}
\renewcommand{\shorttitle}{TripJudge: A Relevance Judgement Test Collection for TripClick Health Retrieval}

\begin{abstract}
Robust test collections are crucial for Information Retrieval research. Recently there is a growing interest in evaluating retrieval systems for domain-specific retrieval tasks, however these tasks often lack a reliable test collection with human-annotated relevance assessments following the Cranfield paradigm.
In the medical domain, the TripClick collection was recently proposed, which contains click log data from the Trip search engine and includes two click-based test sets.
However the clicks are biased to the retrieval model used, which remains unknown, and a previous study shows that the test sets have a low judgement coverage for the Top-10 results of lexical and neural retrieval models.
In this paper we present the novel, relevance judgement test collection TripJudge for TripClick health retrieval. We collect relevance judgements in an annotation campaign and ensure the quality and reusability of TripJudge by a variety of ranking methods for pool creation, by multiple judgements per query-document pair and by an at least moderate inter-annotator agreement.
We compare system evaluation with TripJudge and TripClick and find that that click and judgement-based evaluation can lead to substantially different system rankings.
\end{abstract}



\maketitle

\section{Introduction}

Reliable and robust evaluation of ranking systems is crucial to Information Retrieval (IR) research. Thus a great effort in IR research is put into creating reusable and robust test collections \cite{zobel1998reliability,vorhees2018reusabletest,vorhees2022oldtrec,soboroff2017buildingtest} in task-specific evaluation campaigns like TREC \cite{clarke2012trec,craswell2021trecdl} or CLEF. These campaigns follow the Cranfield paradigm \cite{cyril1991cranfield} to create relevance judgements on the pooled output of the participating systems.
Recently there has been a growing interest in evaluating the retrieval performance of retrieval models for domain-specific retrieval tasks \cite{thakur2021beir,jingtao2022extrapolationdr,hofstaetter2022tripclick,Hofstaetter2022colberter,althammer2021crossdomain} including the medical domain \cite{rekabsaz2021tripclick,kirk2017precisionmedicine,xiong2020cmttreccovid}. Domain-specific retrieval tasks often lack a reliable test collection with human relevance judgments following the Cranfield paradigm \cite{rekabsaz2021tripclick,thakur2021beir}. Furthermore it remains unclear how well old test collections can be used to evaluate neural retrieval models, which were not part of the pooling process \cite{vorhees2022oldtrec}.

In the medical domain, a recently proposed benchmark is the publicly available TripClick collection \cite{rekabsaz2021tripclick}, which contains large-scale click logs from Trip, an English health search engine with professional and non-professional users. It provides two test sets with labels based on the clicks of the users, either estimating relevance by the raw clicks (`Raw') or by the rate of clicks of a document over all retrieved documents for a query (`DCTR').
As the TripClick test sets are based on the clicks of the users, the test sets are biased towards the retrieval model employed by the search engine \cite{ryen2013beliefsbiases}, which remains unknown. In a previous study \cite{hofstaetter2022tripclick} the test sets were shown to have a low annotation coverage of at most $41\%$ of the Top10 results for lexical and neural retrieval models.

In this paper we address these shortcomings of click-based test collections by creating TripJudge, a relevance judgement test collection for TripClick. We collect relevance judgements by running an annotation campaign on the test set queries of TripClick. In order to increase the reusability of our test collection \cite{buckley2007biaslimitslargecollection}, we use three participating systems for the pool creation from \citet{hofstaetter2022tripclick} employing lexical and neural retrieval models. To control the quality of the relevance assessments we monitor the annotation time per query, we employ a graded relevance scheme \cite{grady-lease-2010-crowdsourcing,alonso2012usingcrowdsourcingtrec} and we employ multiple relevance assessments per query-document pair (we aim for three assessments but have at least two). We reach an at least moderate inter-annotator agreement measured with Cohen's Kappa.

Furthermore we compare the click and judgement-based evaluation of various retrieval systems and investigate how the rankings of the systems change. Related work about comparing clicks and judgements for evaluation come to different conclusions. While \citet{joachims2005clickthrough} and \citet{zobel1998reliability} find reasonable agreement between the clicks and relevance judgements, \citet{kamps2009comparative} demonstrate that system rank correlations between evaluation based on clicks or judgements is low. First we analyze the overlap of the click-based test collections with TripJudge and find that the majority of the documents that are judged as relevant are not labelled in the click-based collections and therefore are considered irrelevant during evaluation.
Similar to related work \cite{vorhees1998kendallstau,buckley2004incompleteinfo,vorhees2002effect}, we measure system rank correlation between evaluation with different test collections with Kendalls $\tau$ correlation \cite{kendall1938measure}. We find that the rankings with the evaluation of TripJudge differ from the rankings with the click-based test collections.
%
%
Our contributions are the following:
\begin{itemize}
    \item We create the relevance judgement-based test collection TripJudge for TripClick health retrieval and make it publicly available on \url{www.github.com/sophiaalthammer/tripjudge};
    \item We ensure the quality and reusability of TripJudge by a variety of systems for pool creation, by multiple judgements per query-document pair, and by an at least moderate inter-annotator agreement in our annotation campaign;
    \item We compare evaluation with click-based TripClick and our judgment-based TripJudge and find that click and judgment-based evaluation can lead to different system rankings
\end{itemize}

\section{Methodology}

We describe how we preprocess the TripClick test queries as well as the pool creation followed by the annotation campaign.

\subsection{Data and Pool Preparation}

In the TripClick test sets the queries are grouped by their user interaction frequency into Head, Torso and Tail. For the annotation campaign we used the $1175$ head queries, which consist of keyword queries. During the campaign we noticed duplicate queries which differ in their casing (lower/uppercasing) or queries without any text (for example "\#1 or \#2"). We discard these queries from our TripJudge test collection and end up with $1136$ unique queries\footnote{We publish the reasons for removal, the removed, and remaining queries in GitHub}.

For the pool creation we use the runs from \citet{hofstaetter2022tripclick}. In order to have different first stage retrieval methods we use the lexical retrieval run with BM25 \cite{bm25} (run 1 in Table \ref{tab:evalresults}) as well as the SciBERT$_{DOT}$ run (run 2 in Table \ref{tab:evalresults}) which is based on dense retrieval \cite{Hofstaetter2021_tasb_dense_retrieval,althammer2022parm}. As additional run we use the Ensemble which re-ranks BM25 Top-200 candidates using an Ensemble of BERT$_{CAT}$ based on SciBERT, PubMedBERT-Abstract and PubMedBert-Full Text (run 7 in Table \ref{tab:evalresults}). We create the pool by taking the union of the query-document pairs from the Top-10 of the three runs for all test queries and keep the highest rank among the three runs. This results in a total of $29581$ pairs and prioritize the annotation according to the rank of the document: all Top-n pairs have priority $10-k (k \in [1..n])$; the higher the priority, the earlier they are selected for annotation in the annotation interface.

In order to maintain a low latency in our annotation system, we needed to truncate the documents. As the dense retrieval models rely on the text up to 512 BERT tokens, we truncate the document text to this length, which applies to $10\%$ of the documents.

\subsection{Annotation Campaign}

We conduct the annotation campaign among $135$ computer science students with a target of $300$ annotations per annotator and reach an average of $287$ annotations per annotator (Table \ref{table:statisticscampaign}).

The users of the Trip search engine are a mix of experts and non-experts. As the annotators are non-experts and as previous work points out the challenges with students as annotators \cite{palotti2016studentsasssessors,bailey2008judgesexchangable}, we take several steps to ensure and monitor the quality of the annotations:
\textbf{1)} We use a 4-graded, ordinal relevance scheme: \textit{Wrong (1)}, \textit{Topic (2)}, \textit{Partial (3)}, and \textit{Perfect (4)}, as suggested in related work \cite{grady-lease-2010-crowdsourcing,alonso2012usingcrowdsourcingtrec}
\textbf{2)} We aim for three relevance assessments per query-document pair, on average we reach $2.92$ relevance assessments per pair, where we employ majority voting or a heuristic in case of no majority. We discard the pairs with only one relevance assessment.
\textbf{3)} We monitor the average annotation time per pair per relevance grade, we remove annotations with a short annotation time (below 1 second) and reach an average annotation time of $48$ seconds per judgement.
\textbf{4)} We collect feedback about the campaign from the students at the end of the annotation campaign.

We conducted the annotation campaign using the FiRA interface \cite{hofstaetter2020neuralirexplorer,Hofstaetter2020_fira} and ran the campaign for 7 days with a fixed deadline\footnote{We publish the annotation guidelines in our GitHub repository}. 
To control the quality of the judgements during the campaign, we monitored the average number of judgements per 12 hours and we observe the daily average annotation time per relevance grade to detect random judgements. 
We reach a high average annotation time per relevance grade (Table \ref{table:statisticscampaign}).

Overall the students gave us positive feedback about the evaluation campaign: $33\%$ rated it \textit{Very good}, $28\%$ \textit{Good} and $27\%$ \textit{Decent}, only $12\%$ of the students did not like it. The students could also give written feedback. The main feedback was that it was hard to distinguish between the relevance grades of  \textit{Topic} and \textit{Partial}. This difficulty is also reflected in the average time per annotation for these relevance grades. While it took the annotators on average $39$ seconds to decide on \textit{Wrong}, the annotations for \textit{Topic} and \textit{Partial} took $47$ and $54$ seconds on average, respectively.
We reached $38810$ total judgements and judged at least until the Top-4 for our pool.

\begin{table}
    \label{tab:datasetstat}
    \setlength\tabcolsep{2pt}
    \begin{tabular}{lr}
       \toprule
       \# of queries & 1136 \\
       \# of documents in collection & 2.3M\\
       \midrule
        \# annotated q-d pairs & 12590\\
        \# of judgements & 38810\\
        avg \# assessments per query & 2.92\\
       \midrule
       \# (\%), avg annotation time for Wrong & 3811 (10\%), 39s\\
       \# (\%), avg annotation time for Topic & 10901 (28\%), 47s\\
       \# (\%), avg annotation time for Partial & 13008 (33\%), 54s\\
       \# (\%), avg annotation time for Perfect & 11090 (29\%), 44s\\
       \midrule
       \# annotators & 135\\
       avg \# of annotations per annotator & 287 \\
        \bottomrule
    \end{tabular}
    \caption{Statistics of the annotation campaign.}
    \label{table:statisticscampaign}
    \vspace{-0.9cm}
\end{table}


\section{Quality analysis}

After the annotation campaign we processed the $38810$ raw relevance judgements to form the TripJudge test collection which we publish in the standard TREC format for qrels.

\begin{figure}
    \centering
    \includegraphics[width=.9\linewidth]{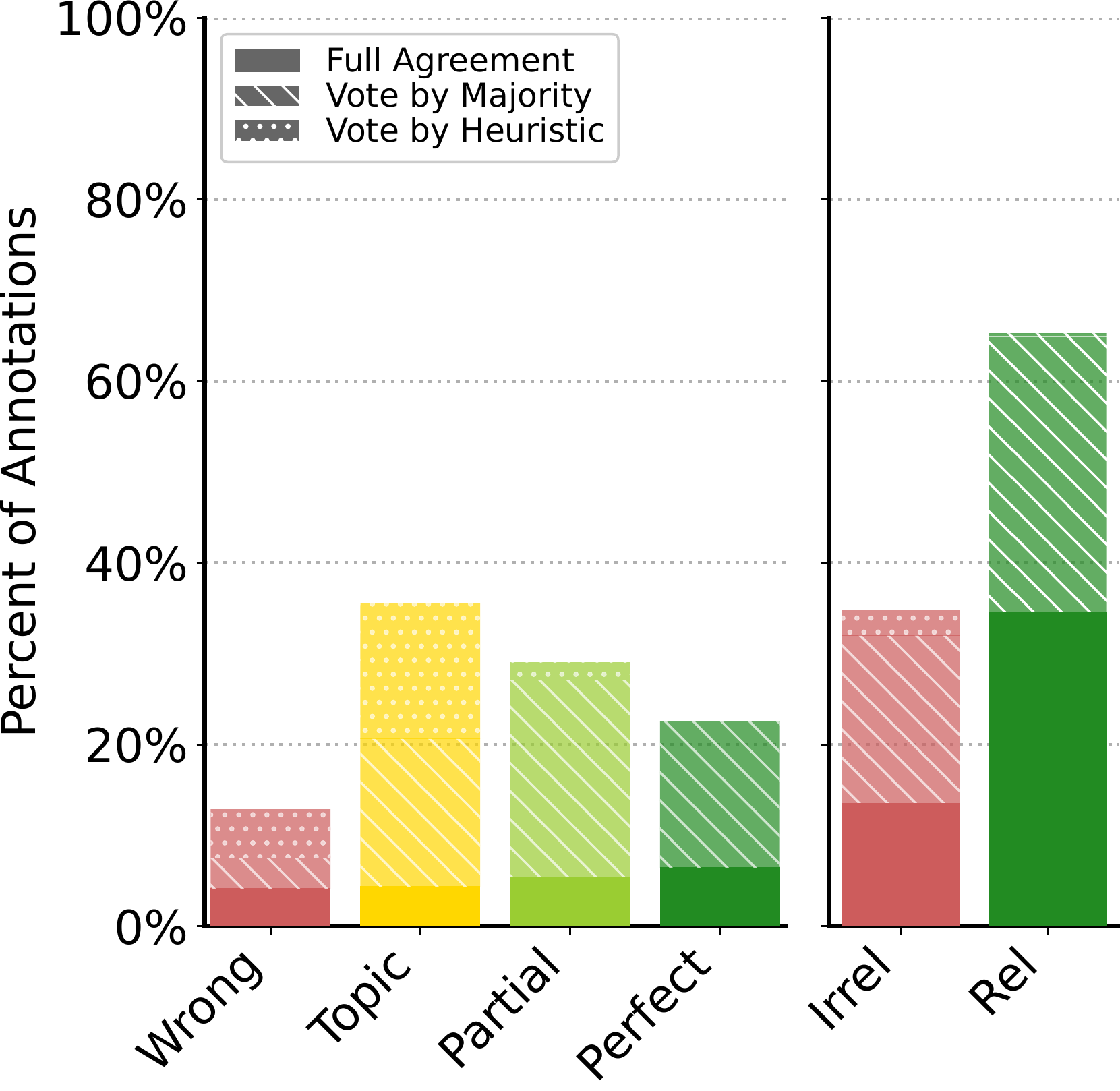}
    \vspace{-0.5cm}
    \caption{Distribution of relevance grades for 4-grade and 2-grades, percentage of heuristic and majority voting.}
    \label{fig:distributionlabels}
\end{figure}

\begin{figure}
    \centering
    \includegraphics[width=0.9\linewidth]{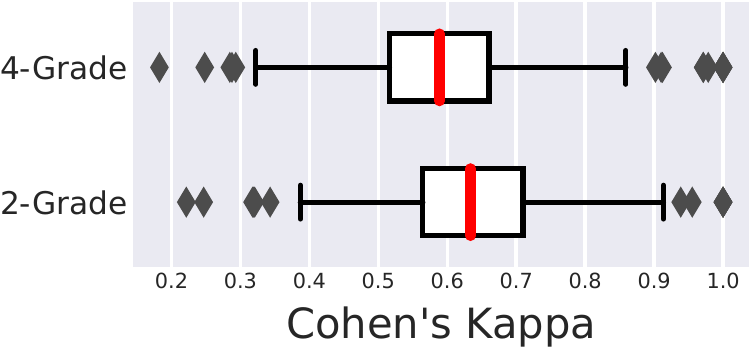}
    \vspace{-0.5cm}
    \caption{Cohen's Kappa agreement between the annotators and the annotations aggregated with majority voting.}
    \label{fig:kappa}
    \vspace{-0.5cm}
\end{figure}

We removed the query-document pairs with only one judgement. We grouped the relevance grades \textit{Wrong} and \textit{Topic} into \textit{Irrelevant} and \textit{Partial} and \textit{Perfect} into \textit{Relevant}, in order to attain a 2-graded relevance judgement with potential higher agreement.
We computed the final relevance judgement either via full agreement, if all annotators agree on the relevance grade or with majority voting, if the annotators disagree, or with the heuristic of taking the lowest relevance grade, if the annotators disagree and there is no majority for a relevance grade. We apply this heuristic with the assumption that if disagreement is high, the relevance cannot be definitely decided and therefore the document should be annotated as irrelevant.
In Figure \ref{fig:distributionlabels} we visualize the 4-grade and 2-grade distribution of the judgements and their percentage of full agreement, majority voting and the heuristic. While the full agreement is low ($20\%$ of all queries) for the 4-grade relevance judgements, the full agreement for the 2-grade relevance judgements is substantially higher with $48\%$ of all queries. Furthermore, a high percentage of judgements is decided via majority vote. For the 4-grade relevance judgements, $22\%$ of the queries are decided by the heuristic which indicates a high disagreement between the annotators, but the percentage of heuristic decisions for the 2-grade relevance judgements is low with $2\%$. This shows that the 2-grade relevance judgements are more robust and have a higher agreement between the annotators.

We also study the inter-annotator agreement between the annotators for all relevance judgements. We measure the inter-annotator agreement with Cohen's Kappa $\kappa$ \cite{cohen1960kappa}, which is a standard metric to compare multiple sets of judgements and to measure the subjectivity in the assessments. As the 4-grade annotations are ordinal, we use a linear weighted Kappa for them. 

\begin{figure}
    \centering
    \includegraphics[width=.98\linewidth]{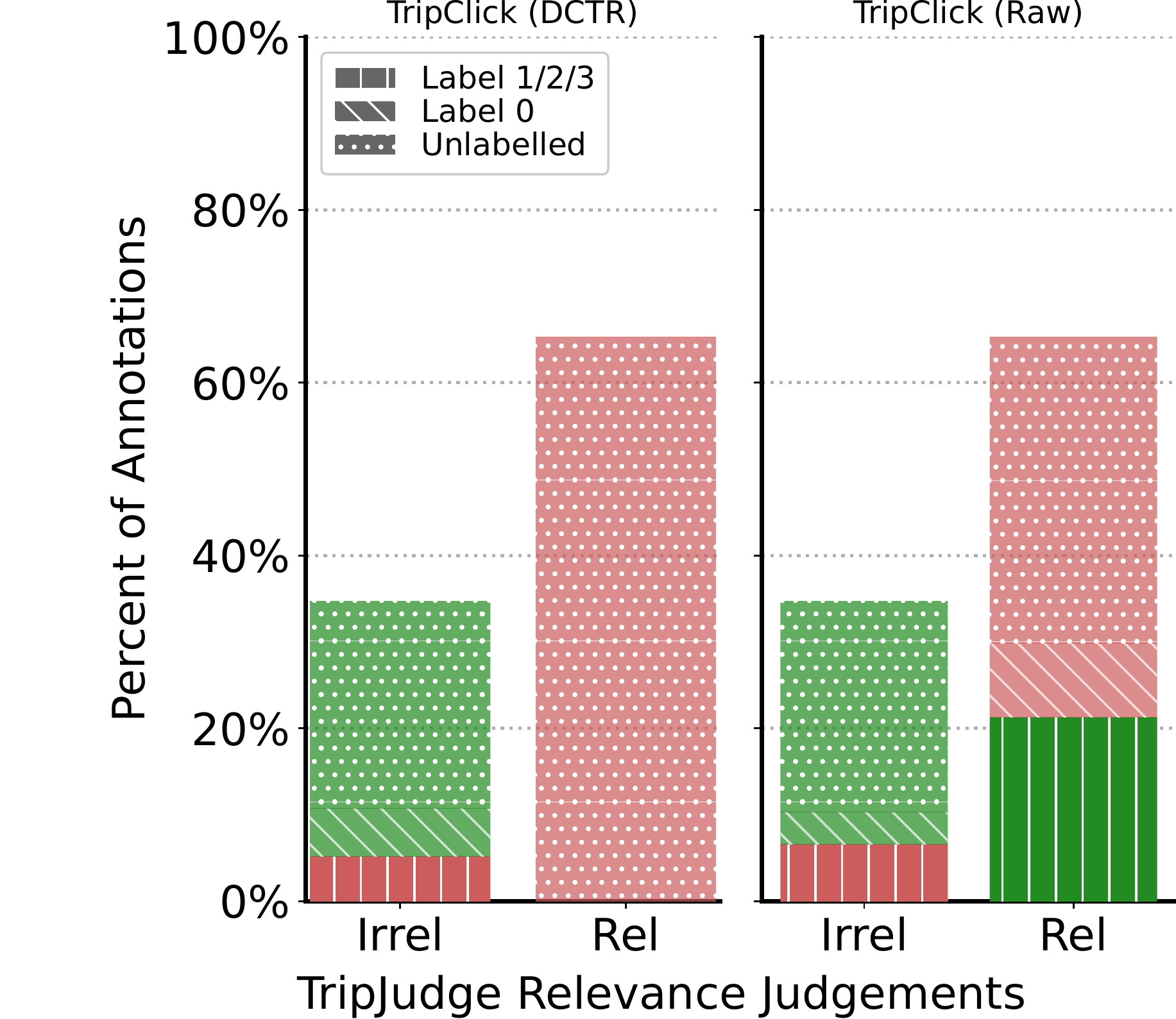}
    \vspace{-0.25cm}
    \caption{Relevance judgements from TripJudge for the Top-4 of the pool vs TripClick click-based labels from the DCTR or Raw test collection. Green bars denote agreement between the relevance judgement of TripJudge and the click label from TripClick, red bars denote disagreement.}
    \label{fig:intersectionlabels}
    \vspace{-0.5cm}
\end{figure}

In Figure \ref{fig:kappa} the 2-grade and 4-grade agreement with Cohen's Kappa is visualized with an average $\kappa$ of $0.63$ for the 2-grade and an average  weighted $\kappa$ of $0.60$ for the 4-grade relevance judgements, which indicates moderate agreement for the 4-grade and substantial agreement for the 2-grade. For both the 4-grade and 2-grade agreement we reach an at least moderate agreement with $50\%$ of the kappa values between $0.50$ and $0.70$.
Furthermore, a certain level of disagreement in the relevance judgement is expected due to the subjectivity of the individual annotators \cite{borlund2003relevanceir,verberne2017automatic} and our agreement levels align with previous work \cite{alonso2012usingcrowdsourcingtrec}, which also employs non-expert annotators for judgements of TREC collections.

\section{TripJudge vs TripClick}

We compare the relevance judgements of TripJudge with the click-based labels of TripClick and investigate the system ranking difference of the two test collections. Due to the higher inter-annotator agreement we consider the 2-graded judgements of TripJudge.


\begin{table*}
    \centering
    \setlength\tabcolsep{3pt}
    \begin{tabular}{cl!{\color{lightgray}\vrule}rr!{\color{lightgray}\vrule}rrrr!{\color{lightgray}\vrule}rr!{\color{lightgray}\vrule}rrr!{\color{lightgray}\vrule}rr!{\color{lightgray}\vrule}rrr}
       \toprule
       \multirow{2}{*}{} & \multirow{2}{*}{\textbf{Model}} & 
       \multicolumn{6}{c!{\color{lightgray}\vrule}}{\textbf{TripJudge}}&
       \multicolumn{5}{c!{\color{lightgray}\vrule}}{\textbf{TripClick (DCTR)}}&
       \multicolumn{5}{c}{\textbf{TripClick (Raw)}}\\
       && \small{J@5}& \small{J@10}&\small{n@5-j}& \small{n@5} & \small{n@10}& \small{R@100}& \small{J@5}  & \small{J@10}
& \small{n@5} & \small{n@10}& \small{R@100}& \small{J@5} & \small{J@10}
&  \small{n@5} & \small{n@10}& \small{R@100}  \\
        \midrule
         \multicolumn{6}{l}{\textbf{First stage retrieval}} \\
         \textcolor{gray}{1} & BM25      &78\%&47\%&\textbf{.761}&.694&.570&.771&33\%& 31\%& .122& .140 &  .499 &30\%&27\%& .199 & .198  & .464  \\
         \textcolor{gray}{2} &      SciBERT$_\text{DOT}$      &87\%&50\%&.602&.540&.456&.636&48\%&41\%& .232&  .243 &  .562 &44\%&38\% &.362&   .328&.496   \\
         \textcolor{gray}{3} &          PubMedBERT$_\text{DOT}$    &49\%&36\%&.652&.377&.356&.649&45\%&40\%&.223&   .235  &  .582 &42\%& 37\% &.345& .318 &.518  \\
        \midrule
         \multicolumn{6}{l}{\textbf{Re-Ranking (BM25 Top-200)}} \\
         \arrayrulecolor{lightgray}
         \textcolor{gray}{4} &ColBERT (SciBERT)             &64\%&44\%&.758&.538&.501&.790&52\%& 47\% &.254 & .270 &.589 &49\%& 43\% &.395& .367  &  .547 \\
         \textcolor{gray}{5} &  ColBERT (PubMedBERT)                       &63\%&44\%&.758&.527&.493&.777&55\%&49\%&.261&  .278  &.595&52\%&45\% & .412 & .382  &.551  \\
         \textcolor{gray}{6} &  BERT$_\text{CAT}$                           &64\%&45\%&.757&.540&.506&\textbf{.818}&56\%&50\% &.271&  .287  & .594 &53\%& 46\% &.421& .389 &.552   \\
         \textcolor{gray}{7} & Ensemble 
         &\textbf{88\%}&\textbf{51\%}&.756&\textbf{.698}&\textbf{.592}&.814&\textbf{58\%}& \textbf{52\%} &\textbf{.285}&  \textbf{.303}  & \textbf{.600} &\textbf{55\%}&\textbf{48\%}  &\textbf{.443}&\textbf{.409} &\textbf{.556}  \\
        \arrayrulecolor{black}
        \bottomrule
    \end{tabular}
    \caption{Effectiveness results and judgement coverage for judgement-based TripJudge and click-based TripClick DCTR/Raw test collection. J@m denotes the judgement coverage at rank m, n@m denotes the nDCG at cutoff m, -j denotes the j-option in trec\textunderscore eval when only evaluating on the judged query-document pairs. Top-10 of run 1,2,7 create the pool for TripJudge.}
    \label{tab:evalresults}
    \vspace{-0.7cm}
\end{table*}

\begin{table}[t!]
    \centering
    \label{tab:datasetstat}
    \setlength\tabcolsep{2pt}
    \begin{tabular}{lrrr}
    \toprule
       Measure & TripJudge--DCTR & TripJudge--Raw & DCTR--Raw\\
       \midrule
       nDCG@5 &0.333 &0.333&1.000\\
       nDCG@10 &0.428 &0.428&1.000\\
       MRR@10 &0.238 &0.238&1.000\\
       Recall@100 &0.238 &0.333&0.714\\
       \bottomrule
    \end{tabular}
    \caption{Kendall tau correlation between system rankings of TripJudge and TripClick DCTR/Raw for four metrics.}
    \label{table:kendallstau}
    \vspace{-0.9cm}
\end{table}

\subsection{Coverage and Intersection}

We analyze the coverage and intersection of the annotated query-document pairs between TripJudge and TripClick DCTR and Raw test collection.
Figure \ref{fig:intersectionlabels} visualizes the percentage of relevant and irrelevant relevance judgements from TripJudge. The different patterns of the bar visualize the label from the TripClick DCTR or Raw test collection. The labels 1/2/3 from TripClick refer to relevant documents, Label 0 denotes irrelevant documents and unlabelled documents are considered as irrelevant during evaluation. The green bars denote agreement between the annotation from TripJudge and TripClick, the red bars denote disagreement. It is striking that all of the relevant documents from the Top-4 of TripJudge are unlabelled in DCTR and therefore considered as irrelevant. Furthermore there is high disagreement between the judgements of TripJudge and click-based labels of TripClick DCTR and Raw.


\subsection{System evaluation}

We compare the system rankings and the coverage of the relevance judgements for the runs for TripJudge and TripClick DCTR and Raw. For TripJudge and TripClick unlabelled documents are considered as irrelevant.
In Table \ref{tab:evalresults} the effectiveness metrics as well as the judgements coverage measured as $J$ at rank $n$ is displayed for various lexical and neural retrieval systems from \citet{hofstaetter2022tripclick}.
For TripJudge we see that the coverage measure with J@5 for the runs in the pool (run 1,2,7) is high (around $80\%$) compared to the coverage of the runs which did not participate in the pooling. However the coverage of TripJudge is higher than the coverage of TripClick DCTR and Raw for the respective runs.
The Ensemble of BERT$_{CAT}$ based on SciBERT, PubMedBERT-Abstract and PubMedBert-Full Text (run 7) consistently reaches the highest retrieval performance for the judgement and click-based test collections. Interestingly the dense retrieval model SciBERT$_{DOT}$ (run 2) underperforms BM25 (run 1) when evaluated with TripJudge, although showing substantially higher retrieval effectiveness for the click-based test collections. This suggests that the higher retrieval effectiveness of run 2 compared to run 1 for the click-based collections is due to the higher coverage of annotations.

Furthermore, we compare the difference in system rankings between two test collection with Kendall $\tau$ \cite{kendall1938measure}, which is a common measure to compare the correlation between two system rankings \cite{vorhees1998kendallstau,sanderson2007problemstau}. In Table \ref{table:kendallstau} are the Kendall tau correlations between two rankings of two test collections regarding four metrics. Test collections with $\tau>0.9$ are considered equivalent \cite{vorhees1998kendallstau}.

For the comparison of TripJudge with the click-based test collections, we see a low correlation of the system rankings for all $4$ evaluation metrics. For the click-based test collections we see that they are equivalent for most of the metrics. We conclude that the system rankings differ drastically between TripJudge and TripClick and that TripJudge offers a valuable and reusable relevance judgement set beside the TripClick test sets for evaluating retrieval systems.

\section{Conclusion and future work}

We present the TripJudge test collection with $38810$ relevance judgements for TripClick health retrieval. We describe the annotation campaign for creating the relevance judgements. For increased reusability we used lexical and neural retrieval systems for pool creation. We 
reach an at least moderate inter-annotator agreement. When comparing the relevance judgements of TripJudge with the click-based annotations from the TripClick test collections, we find that a majority of judged relevant documents were unlabelled in TripClick, and there is a high disagreement between the relevance judgements and the click-based annotations.
We re-evaluate lexical and neural models and find a higher judgement coverage for the retrieval runs for TripJudge than for the TripClick test collections.
The system rankings substantially differ between the evaluation with the relevance-based and click-based collections. 

One limitation of TripJudge is the depth of the relevance judgements as usually a half \cite{craswell2021trecdl} or a third of the annotated query-document pairs should be relevant \cite{vorhees2018reusabletest,vorhees2021qualitycovidir} while we have 65\% relevant. Therefore we see possible future work in annotating to a higher depth. Nevertheless we view TripJudge as a valuable resource for evaluation of the domain specific task of health retrieval.
In conclusion, we argue that there must be more effort put into creating relevance judgements based test collections for domain specific retrieval tasks, in order to evaluate different systems in a robust and conclusive way.

\section*{Acknowledgements}
This work was supported by EU Horizon 2020 ITN/ETN on Domain Specific Systems for Information Extraction and Retrieval(ID:860721).

\bibliographystyle{ACM-Reference-Format}
\balance
\bibliography{sample-base}

\end{document}